\documentclass[a4paper,12pt]{amsart}
\usepackage{geometry}
\usepackage{setspace}
\usepackage{amsmath,amssymb,amsthm,enumitem}
\usepackage{setspace}
\usepackage{amsmath}
\usepackage{graphicx}
\usepackage[]{color} 
\usepackage{epsfig}
\usepackage{latexsym}
\usepackage{amsmath}
\usepackage{amsthm}
\usepackage{graphicx}
\usepackage{amssymb}
\usepackage{hyperref}
\usepackage{tablefootnote} 
\usepackage{etoolbox} 
\usepackage{lipsum}
\usepackage{adjustbox}
\AtBeginEnvironment{tabular}{\doublespacing}

\usepackage{setspace}

\theoremstyle{definition}

\newcommand{\cE}{\mathcal{E}}

\newcommand{\PP}{\mathbb{P}}
\newcommand{\EE}{\mathbb{E}}

\newcommand{\G}{\mathcal{G}}

\newcommand{\cR}{\mathcal{R}}
\newcommand{\cQ}{\mathcal{Q}}

\newcommand{\VV}{\mathbb{V}}
\newcommand{\blue}{\textcolor{black}}

\newcommand{\citep}{\cite}
\newcommand{\citeyear}{\cite}
\newcommand{\citeauthor}{\cite}
\doublespace

\title{An Evolutionary Process without Variation and Selection}

\author{Liane Gabora and Mike Steel}

\begin{document}

\noindent{Full reference: \\Gabora, L. \& Steel, M. (2021). An evolutionary process without variation and selection. {\it Journal of the Royal Society Interface, 18}(180). 20210334. DOI: https://doi.org/10.1098/rsif.2021.0334}\\[1cm] 

\begin{singlespace}
\begin{abstract} 
Natural selection successfully explains how organisms accumulate adaptive change despite that traits acquired over a lifetime are eliminated at the end of each generation. However, in some domains that exhibit cumulative, adaptive change---e.g., cultural evolution, and earliest life---acquired traits are retained; these domains do not face the problem that Darwin's theory was designed to solve. Lack of transmission of acquired traits occurs when germ cells are protected from environmental change, due to a self-assembly code used in two distinct ways: (i) actively interpreted during development to generate a soma, and (ii) passively copied without interpretation during reproduction to generate germ cells. Early life and cultural evolution appear not to involve a self-assembly code used in these two ways. We suggest that cumulative, adaptive change in these domains is due to a lower-fidelity evolutionary process, and model it using Reflexively Autocatalytic and Foodset-generated networks. We refer to this more primitive evolutionary process as Self--Other Reorganisation (SOR) because it involves internal self-organising and self-maintaining processes within entities, as well as interaction between entities. SOR encompasses learning but in general operates across groups. We discuss the relationship between SOR and Lamarckism, and illustrate a special case of SOR without variation.
\end{abstract}
\end{singlespace}

\maketitle


\smallskip

\singlespace{
\noindent Addresses for Correspondence: \\

\smallskip
\noindent Liane Gabora \\
\noindent Department of Psychology, University of British Columbia, Kelowna BC, Canada \\ 
\noindent \email{liane.gabora@ubc.ca}\\

\smallskip
\noindent Mike Steel\\
\noindent Biomathematics Research Centre, University of Canterbury, Christchurch, New Zealand \\
\noindent \email{mike.steel@canterbury.ac.nz}
}

\newpage
\doublespace

\section{Introduction}

Evolution refers to cumulative, adaptive change over time; thus, an evolutionary process produces increasingly adapted entities.
Under standard models of natural selection (dating back to Darwin), evolution is driven by differential replication of randomly generated heritable variations in a population over generations, resulting in some traits becoming more prevalent than others.\footnote{The term `selection’ is used in an informal sense in a way that is synonymous with `choosing’ or `picking out’. One could say that selection—-in the everyday sense of the term-—occurs in a competitive marketplace through the winnowing out of inferior products. However, the term `selection' is used here in its scientific sense.}
Natural selection has provided the foundation of modern biology, and a unifying principle for describing species change well after the origin of life. However, as noted by Gould \citeyear{gou87}, ``Evolution, in fact, is not the study of origins at all... Evolution studies the pathways and mechanisms of organic change following the origin of life.''
Since natural selection requires the existence of multiple competing entities that are selected amongst, it cannot explain how the first living entity arose (nor the second) \citep{gab06}.
The theory of natural selection is an explanation for how life diversified after it originated; the means by which it originated is currently an unsolved problem \citep{ricetal}. 

Darwin's theory arose in response to the paradox of how organisms accumulate adaptive change despite that traits acquired over a lifetime are eliminated at the end of each generation. He devised a population-level explanation: although acquired traits are discarded, inherited traits are retained, so evolution is due to preferential selection for those inherited traits that confer fitness benefits on their bearers.
But although Darwinian principles have been applied to early life \cite{cor15, fer07, fer08} and to cultural evolution \cite{boyric85,cavfel81,erm21,mes04,ste18,whi19}, neither origin of life research, nor cultural evolution research, is plagued by the problem that acquired traits are extinguished at the end of a generation. 
By culture, we mean extrasomatic adaptations such as behavior and artifacts that are socially rather than sexually transmitted, as well as the {\it complex whole} of the knowledge, beliefs, art, laws, customs, and so forth, acquired by the members of a society \citep{tyl71}.
Because of lateral or horizontal transfer in both culture and very early life, these domains do not face the paradox that Darwin's theory was designed to solve \cite{gab06,gab11,vet06}.

One could argue that even in these domains where acquired traits are not eliminated at the end of each generation, the observed cumulative, adaptive change may nevertheless be due to natural selection. However, species that discard traits acquired over a lifetime possess a set of coded self-assembly instructions which are used in two distinct and separate ways: (i) actively interpreted during development to generate a soma, and (ii) passively copied without interpretation during reproduction to produce germ cells \cite{von66,lan92}. This sequestering of germ cells from developmental changes is responsible for the \emph{sine qua non} of a Darwinian process: lack of transmission of acquired traits \cite{hol92}. This tells us that it is a rather complex structure---i.e., a set of coded self-assembly instructions used in these two distinct ways---that enables cumulative, adaptive change without transmission of acquired traits. The above-mentioned forms of evolution (that of earliest life, and cultural evolution) may not possess such a structure. \blue{Note also that although culture involves replication, as when one artist copies a painting made by another, this is not {\it self}-replication; the painting does not replicate itself.}

Accordingly, it would seem appropriate to pay particular attention in these domains to the role of epigenetic and non-Darwinian (e.g., Lamarckian) processes, and indeed, it is increasingly accepted that non-Darwinian processes play a role in evolution \cite{gab06,kau1,kil19,vet06,woe02}.
This paper goes further by providing an existence proof that evolution is possible in the {\it absence} of variation and selection, and mathematically describing a type of entity that allows for this kind of evolutionary process. This second component is important, because although entities of many kinds change due to acquired traits---e.g., a rock tumbling down a stream will acquire rounded edges---such change may not be adaptive. 
In contrast to the rock---and also, in contrast to neutral evolution \cite{kim68,kin69}---the evolutionary change considered in this paper is adaptive, i.e., it contributes to the entity's efficiency, survival, or replication. 
Humans might find smoother rocks more appealing, and thus they may be more valuable to humans, but this increased smoothness is not valuable to the rock itself; indeed, the process that rounds the edges of a rock culminates in its disintegration. 
Similarly, optimization algorithms produce outcomes that are deemed adaptive not because they benefit the algorithms themselves; they are deemed adaptive only insofar as they benefit us.


In origin of life research, there is a sizeable literature on models of the evolvability of primitive metabolic networks prior to DNA (e.g., \cite{lan18, seg00, she07, vas12, xav20}). Here, in order to describe the origin of both
biological life and culture, within a common framework for the origins of evolutionary processes, we take a more general mathematical approach. 
The approach addresses the problem of network control \cite{lil19} using a generalized notion of catalysis, or facilitated interaction; existing elements (referred to as the \emph{foodset}) interact, and stimulate (or catalyze) the generation of new (foodset-\emph{derived}) elements, thereby introducing a means of endogenous control.

The paper begins with an introduction to the general framework for modeling the origins of evolutionary processes, which will be used to develop the argument. We then provide a mathematical model for evolution in the absence of variation and selection with this framework.
Next, 
we explain why we refer to this form of evolution as Self--Other Reorganisation (SOR), and describe what type of structure is able to exhibit this form of evolution.
To accommodate an interdisciplinary readership, a list of standard definitions of terms used in this paper is provided in the Supplementary Materials. Acronyms used in this paper are listed alphabetically in Table ~\ref{acro}.

\bigskip

\begin{table}[ht]
\begin{center}
\begin{tabular}{@{} lll @{}}
\hline \hline 
\textbf{Acronym} & \textbf{Meaning}\\ 
\hline
CRS &  Catalytic Reaction System \\
RAF & Reflexively Autocatalytic and Foodset-generated (F-generated)\\
OOL   & Origin of Life \\
OOC  & Origin of Culture \\
MR & Mental Representation \\
CCP & Cognitive Catalytic Process \\
SOR  & Self--Other Reorganisation \\
\hline
\hline
\end{tabular}
\end{center}
\caption{Acronyms used in this paper.}
\label{acro}
\end{table}

\section{Autocatalytic networks}

The evolution process described here involves entities that (1) are self-maintaining (i.e., they have a  
means of 
preserving their structure), and (2) interact with each other by way of their environment.
The theory of autocatalytic networks grew out of studies of the statistical properties of {\it random graphs} consisting of nodes randomly connected by edges \citep{erdosrenyi1960}. As the ratio of edges to nodes increases, the size of the largest cluster increases, and the probability of a phase transition resulting in a single giant connected cluster also increases.  
The recognition that connected graphs exhibit phase transitions led to their application to efforts to model of the origin of life (OOL) \cite{kau2, kau1}.
Applying graph theory to the OOL, the nodes represent catalytic molecules, and the edges represent reactions. 
It is exceedingly improbable that any catalytic molecule present in Earth's early atmosphere catalysed its own formation. However, reactions generate new molecules that catalyse new reactions, and as the variety of molecules increases, the  variety of reactions increases faster. As the ratio of reactions to molecules increases, the probability increases that the system undergoes a phase transition. When, for each molecule, there is a catalytic pathway to its formation, they are collectively {\it autocatalytic}, and the process by which this is achieved has been referred to as {\it autocatalytic closure} \citep{kau1}.
The molecules thereby become a self-sustaining, self-replicating,  living protocell \citep{hord15}. Thus, the theory of autocatalytic networks provides a promising avenue for 
understanding how biological life began \citep{xav20}. 

\subsection{Reflexively Autocatalytic and Foodset-generated (RAF) networks}

Autocatalytic networks have been developed mathematically in the theory of Reflexively Autocatalytic and Foodset-generated (RAF) networks \cite{hor16,rav20,ste19}.
The term {\it reflexively} is used in its mathematical sense, meaning that every element is related to the whole. 
The term {\it foodset} refers to the elements that are initially present, as opposed to those 
produced through interactions amongst them. 
(Elements of a ‘foodset’ are not the same thing as `food’; the foodset is simply the raw building blocks available.)

We now summarise the key concepts of RAF theory. 
A {\it catalytic reaction system} (CRS) is a tuple $\cQ=(X,\cR, C, F)$ consisting of a set $X$ of element types, a set $\cR$ of reactions, a catalysis set $C$ indicating which element types catalyse which reactions, and a subset $F$ of $X$ called the foodset. 
A {\it Reflexively Autocatalytic and F-generated}  set (i.e., a RAF) is a non-empty subset $\cR' \subseteq \cR$ of reactions that satisfies the following properties:

\begin{enumerate}
  \item {\it Reflexively autocatalytic (RA)}: each reaction $r \in \cR'$ is catalysed by at least one element type that is either produced by $\cR'$ or present in the foodset $F$; and
  \item {\it F-generated}: all reactants in $\cR'$ can be generated from the foodset $F$ by using a series of reactions from $\cR'$ itself.
\end{enumerate}

A set of reactions that forms a RAF is simultaneously  self-sustaining (by the $F$-generated condition) and  (collectively) autocatalytic (by the RA condition) because each of its reactions is catalysed by an element associated with the RAF. Note that in RAF theory, a catalyst does not necessarily cause more of something to be produced (or at a faster rate); the catalyst is simply the impetus that allows a reaction to proceed.

A CRS need not have a RAF; alternatively, it may contain many RAFs. When a CRS does contain RAFs there is a unique maximal one, 
the {\it maxRAF}.
It is the fact that a CRS may contain multiple RAFs that allows RAFs to evolve, as demonstrated both in theory and in simulation studies, through selective proliferation and drift acting on the RAFs  that are subsets of the maxRAF \citep{hor16,vas12}.

\subsection{Biological Applications of RAFs}

RAFs have proven useful for modelling the OOL and the onset of biological evolution \cite{hor16,ste19,vas12,xav20}. In the OOL context, a RAF emerges in systems of polymers (molecular strings composed of repeated units called monomers) when the complexity of these polymers (as measured by their maximum length) reaches a certain threshold \cite{kau1,mos}. The phase transition from no RAF to a RAF incorporating most or all of the elements depends on (1) the probability of any one polymer catalyzing the reaction by which a given other polymer is formed, and (2) the maximum length (number of monomers) of polymers in the system. 
This transition has been formalised and analysed mathematically, and using simulations, and RAF theory has been applied to real biochemical systems \cite{horetal10, horetal11, horsteel04, hor16, mos}.
The theory has proven useful for identifying how phase transitions might occur and at what parameter values. 

\subsection{Applications of RAFs to Culture and Cognition}


Computational models of cultural evolution (e.g., \cite{gab95,hen04,koletal15}) have described adaptive change in the absence of autocatalytic structure. However, in such models, the cultural outputs are said to become more adapted (or complex) not because they change in ways that enhance the cultural outputs’ {\it own} preservation or evolution, but because they change in way benefit the artificial agents producing them. 
The programmers may define up front what constitutes a superior cultural output, but the models do not incorporate how different cultural outputs interact with agents in ways that affect survival or evolvability. 
As such, these models are not forced to confront the question of what kind of structure can sustain cultural evolution.
In the autocatalytic model presented in this paper, (1) we do consider how the outputs interact with the internal structure of the agents, modeled as RAF networks, and (2) it is not the outputs that are considered to be evolving, but the self-organizing, self-preserving, and essentially autocatalytic minds that generate and make use of these outputs. 

Autocatalytic networks have been used to model cognitive transitions associated with the origin of culture and the capacity for cumulative cultural evolution \citep{gab98,gab98b,gab00,gabste,gabste20,gabste-modern}, and the onset of behavioral modernity (i.e., the capacity to think and act like modern humans) \citep{gabste-modern}; for related approaches see \cite{andtor19,cab13, endetal11, mut18}. Also, the evolution and reproduction of RAFs has been studied in an OOC context \cite{gabste20}. RAFs have also been used to develop a process model of conceptual change (i.e., the process by which a child becomes an active participant in cultural evolution) \citep{gabbec21}.

Whereas the OOL involves networks of chemical reactions, the OOC setting involves conceptual networks of knowledge and memories, and the products and reactants are not catalytic molecules but culturally transmittable {\it mental representations}\footnote{Although we use the term `mental representation,' our model is consistent with the view (common amongst ecological psychologists and in cognition and quantum cognition communities) that what we call mental representations do not `represent' but  instead act as  contextually elicited bridges between mind and world.} (MRs) of knowledge, experiences, ideas, schemas, and scripts. 
Whereas a chemical reaction network may be bounded by a semipermeable lipid membrane, a conceptual network supervenes upon a brain within a body, which thus constitutes a boundary on the cognitive RAF.
The OOL and OOC settings both exhibit an underlying RAF structure. 
Just as reactions between molecules generate new molecules, interactions between MRs generate new MRs, which in turn enable new interactions. 
In a cognitive/cultural context, the term {\it foodset} refers to MRs that are innate, or obtained through social learning or through individual learning from the natural world or human-made artifacts. In contrast, foodset-generated refers to MRs that are generated from scratch by the individual, as a result of mental operations such as induction or concept combination.
Initially they are generated solely through interactions amongst foodset representations, but as the proportion of foodset-generated items increases they come to play a larger role in these interactions. 
There are generally multiple RAFs in the conceptual network of an individual, which may be subsumed by the maxRAF that connects them; though, they may also be disconnected (as when inconsistent views are held by the same individual). 

The RAF approach to culture and its cognitive underpinnings offers several advantages over other approaches:

\begin{itemize}

\item {\bf Semantic grounding.} the problematic 
circularity of defining MRs in terms of other MRs \citep{har90} is avoided by distinguishing between foodset items and foodset-derived elements. This provides a natural means of grounding abstract concepts in direct experiences.
Complex understandings, modeled as foodset-derived MRs, emerge through mental operations, modeled as `reactions,’ that can be traced back to foodset MRs in a given individual.

\item {\bf Reactivity of ideas.} In cognitive RAFs, mental representations (MRs) become aligned with needs and desires (e.g., the need to solve a problem or resolve cognitive dissonance), and the emotions they elicit. In so doing, they incite (`catalyze’) interactions between other mental representations (modeled as `reactions’), which can be carried out recursively. As a result, whereas conventional conceptual network models require external input to continue processing, RAFs `catalyze’ conceptual change endogenously (in the absence of new external input), resulting in new conduits by which needs and goals can be met. Thus, a RAF network is self-organizing, and conceptual restructuring can percolate throughout the network and affect its global structure. 

\item {\bf Track cultural change within and across conceptual networks.} 
In a RAF model, acquired traits are modeled as foodset-generated elements: traits that come into existence within a given entity. For example, if Alice comes up with an idea herself, for Alice that idea is an acquired trait, modeled as a foodset-generated item. If Alice expresses the idea to Bob, we have transmission of an acquired trait. The idea becomes part of Bob’s foodset, because he did not generate it from scratch. However, if he used Alice’s idea to develop his own new idea, this new idea would be an acquired trait for Bob, modeled as a foodset-generated element.
Thus, the distinction between foodset and foodset-generated elements
makes it possible to trace innovations back to the individuals that generated them, and and track how new ideas and cultural outputs emerge from previous ones.

\end{itemize}


\section{A RAF model of evolution without variation and selection}
\label{varRAF}

\begin{figure}[h!]
\centering
\includegraphics[scale=1]{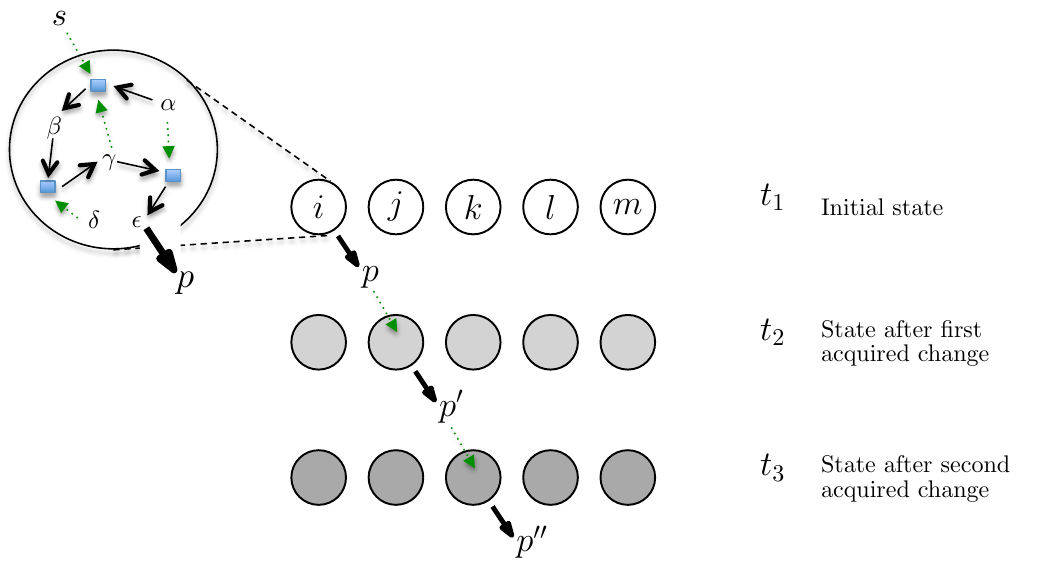}
 \caption{Identical entities exhibit cumulative adaptive change over time, as indicated by transition from light to dark, as a result of cumulative environmental changes, indicated by $p$, $p' \ldots$. 
 \blue{Catalysis is indicated by dashed green arrows. Reactions are indicated by blue squares. Reactants and products within the RAF of entity $i$ are indicated by Greek letters.
 Stimulus $s$ catalyses a reaction in the RAF of entity $i$, resulting in product $p$.} The structure of $i$'s RAF is illustrated in the blow up on the upper left (with the set of element types $X=\{\alpha, \beta, \gamma, \delta, \epsilon, s\}$, food set $F=\{\alpha, \delta, s\}$, and with $\epsilon$ giving rise to $p$). 
 The product $p$ in turn catalyses a reaction sequence in $j$ at time $t_2$, leading to $p'$, and so on. In the scenario outlined here, stimulation of $i$ by $s$ occurs only once, but that is sufficient to generate $p$, which is sufficient to set in motion a sequence of adaptive changes.}
\label{fig1}
\end{figure}

We now demonstrate a primitive non-Darwinian form of evolution using RAF networks. 
Consider a group $\G$ of indistinguishable entities, as illustrated in Fig.~\ref{fig1}. (They may be CRSs such as those associated with very early life \cite{bau18,corcar17,gab06,goletal17,horetal18,ste00,vet06}, culturally evolving human conceptual networks (e.g., \cite{gab95,gabtse17}),  or some other structure we have never encountered.) The entities are described as identical RAFs.

One entity in $\G$, which we call entity $i$, encounters an environmental stimulus, which we refer to as $s$. In the case of a biochemical reaction network, $s$ could be a molecule that crosses the lipid membrane encasing it.
In the case of culture, $s$ could be a visual or auditory stimulus. What makes $s$ a novel stimulus is that at $t_0$ it was not part of the reaction network of any entity in $\G$. Stimulus $s$ is nonetheless part of the foodset, because it is available as an initial reactant or catalyst, as opposed to being directly or indirectly {\it constructed from} available reactants.\footnote{It is a matter of time before one of the entities encountered $s$; the more abundant $s$ is, the sooner this chance encounter. The encounter with $s$ may result from movement of $s$ into the environment, or discovery of a previously uncharted part of the environment rich in $s$.
}

Stimulus $s$ triggers one or more reaction(s) within the RAF network of entity $i$. In a OOL setting the reactions are chemical reactions, while in a OOC setting they are mental operations 
that modify MRs. (Recursive sequences of such `reactions' have been referred to as {\it cognitive catalytic processes} (CCPs) \cite{gabste,gabste20}).
The formation of a transient CCP is illustrated in Fig. 2(ii) of \cite{gabtse17}.)
This reaction (or reaction chain) culminates in the generation of a product $p$ into the environment. Thus, $p$ was not part of the foodset; it is the end result of the series of reactions (or interactions) triggered by $s$. Note that $s$ can only trigger formation of $p$ if the necessary reaction sequence takes place, and 
this reaction sequence requires the RAF. 

Product $p$ confers some adaptive benefit on $i$; exposure to $p$ may improve the efficiency of $i$'s reaction network (or conceptual network) by reducing the number of steps required for a particular reaction sequence (or problem-solving process). It may have a protective function (e.g., by making the environment less inviting for a predator) or (in a cultural context) it may serve as a source of pleasure that enhances wellbeing (e.g., a work of art). 
Having generated $p$, $i$ returns to its initial state, except that it now experiences the adaptive benefit of $p$.

Once $p$ is released into the environment, it can be encountered by other entities in $\G$. (In an OOL context, $p$ may be transmitted because it passes through a lipid membrane or protocell wall. In an OOC context, $p$ may be transmitted via social learning, or by way of an artifact.) 
A recipient of $p$ could have had the capacity to generate $p$ on its own, but by chance had never done so. (In a OOL context, for example, a particular catalyst may have never come in contact with its reactant. In a cultural context, an individual may possess two concepts but never thought of combining them.) Alternatively, it is possible that the recipient did not previously possess the capacity to generate the product.
 
Since all entities are identical to $i$ and exist in the same environment as $i$, they too undergo the adaptive benefit of $p$, and this benefit percolates throughout $\G$.\footnote{The word `percolates’ is used here to refer to the `communal exchange’ (between-entities) component of the process, as opposed to the `self-organization’ (within-entities) component of the process. 
If one entity encounters $p$ before another, that first entity may temporarily function more efficiently.
We note also that the scenario could unfold such that each entity is affected solely by its own products, and thus, after the impact of the initial generation of $p$ by $i$, there could be little interaction, direct or indirect, amongst them.
In other words, it is possible that $j$ is affected by the $p'$ it generated itself (as when someone is affected by looking at their own art). However, it is also possible that $j$ is affected by the $p'$ generated by another entity in $\G$ (as in when someone is affected by looking at art made by someone else). 
Finally, we note that possession of a reactant (which in the cognitive scenario is a MR of a particular product) does not guarantee catalysis of that reactant. For example, someone may have a memory of an artwork, but while the initial experience of it yielded little cognitive restructuring, the second experience of it triggers a productive stream of thought. 
} 
The generation of $p$ makes it possible to generate a new product, $p'$, which confers even greater adaptive benefit than $p$. 
An entity $j$ assimilates $p$, which triggers a sequence of one or more reactions in $j$, resulting in $p'$.  
Entity $j$ benefits from the enhanced adaptive quality of $p'$ over $p$. The ability to produce $p'$ and experience its adaptive benefit percolates throughout the group $\G$.
Thus, each version of $p$ paves the way for its successor. 
The entities in $\G$ continue to generate increasingly beneficial versions of the original $p$ (e.g., $p', p'', p''', \ldots$) catalyzed by the products of other entities. They collectively come to possess the ability to generate and benefit from this family of products through the restructuring of their RAFs.  Thus, we have cumulative adaptive change over time.  Notice that in this model there is no selection and no competitive exclusion, and nor is there birth or death; it is the same set of entities at the beginning as at the end. 

\subsection{Modelling this process}

We consider a simple model based on two processes: percolation and generation. Let $\rho$ denotes the rate of percolation of products, or knowledge (i.e., MRs) of these products through the group $\G$, with neighbourhood structure represented by a directed graph $D_\G$. (The nodes of this graph are the entities in $\G$, and the arcs indicate directed percolation links). 
For the generation process, 
we let $\lambda$ denote the rate (per entity) at which new products are generated. Provided that the ratio $\rho/\lambda$ is not too small, the entities  evolve via cumulative, adaptive change, with the only variation being between those in $\G$ that
have not yet switched to a new  product and those that have (this might be likely in a cultural setting when the entity generating a new product is equidistant from all the others, or in a social media context, where everyone sees something at the same time).  Moreover, for large values of $\rho/\lambda$, each percolation step will be complete before the next new product is generated, so there will be no variation amongst the entities. 

Conversely, as $\rho/\lambda \rightarrow 0$, the entities diverge from each other with respect to the products they generate. Their complexity (which may be estimated through analysis of the maximal number of modifications to any product at any given time for all entities in $\G$) is expected to be lower than in the previous scenario of shared cumulative adaptive change. 

To help formalise the model, we adopt the following terminology.  Given the products  $p, p', p'', \ldots$, we say that the entity that first gives rise to one of these products {\em generates} it.  In the case where this product is produced from an earlier product in the series (rather than being the original product $p$ in response to a stimulus), we say the entity {\em transforms} the existing product to the new one.
We model the increase in the adaptive value of products within $\G$ using two stochastically independent non-deterministic processes. 
 
First, the  generation of new products by entity $i$ involves either generating a new product, or transforming the most recent version of any available product; for simplicity, we assume that these are equally probable. For example, if entity $i$ has the products $(p, p', q, r, r', r'')$ currently available, then it can either transform $p'$, or $q$, or $r''$, or generate a new product, and each of these four outcomes has the same probability, namely $1/4$. This process across entities is assumed to be described by independent exponential random variables with a fixed rate per entity of $\lambda$. (Making $\lambda$ independent of the entity is consistent with the assumption that all entities are initially identical).  
 Second, each newly generated product begins to percolate through the group by moving along the arcs of the directed graph $D_\G$ according to a continuous-time random walk 
 at rate $\rho$.  For simplicity, we will treat  the two processes---generation and percolation---as stochastically independent. 
Let $N=|\G|$ (the number of entities in $\G$). We will assume that the directed graph $D_\G$ that describes the community interactions within $\G$ has the property that from each entity in $\G$, it is possible to reach any other entity in $\G$ by following some directed path in $D_\G$ (i.e., $D_\G$ is `strongly connected');  however, no further assumptions are made regarding this graph.  

We now introduce some notation to keep track of the different versions of products that arise in the process described in Section~\ref{varRAF}. Suppose that  product $p$ first arises from entity $i_1$, and product $p$ is then further modified by entity $i_2$, and so on, with the last modification (before time $t$) being made by entity $i_k$.  
We denote the sequence of products thus generated within 
$\G$ up to time $t$  as: $p(i_1), p(i_1, i_2), p(i_1, i_2, i_3), \ldots$.  More generally, we denote such  a sequence by writing  $(p(i_1), \ldots, p(i_1, i_2, \ldots, i_k): k \geq 1)$ (thereby allowing the possibility that a product is generated but not transformed, in the case where $k=1$).  We refer to the number $k$ of terms in this sequence as the {\em complexity} of the final product;  thus, when an entity transforms a product, it increases its complexity by $1$ (in particular, the first product $p(i_1)$ has complexity 1). 

Note that under the assumptions of the model, the entities $i_1, \ldots, i_k$ are not 
necessarily 
distinct (i.e., an entity may enhance a product more than once,  either consecutively, or later in the sequence). There may also be several such sequences generated within $\G$;  for example, in addition to the previous sequence, one might also have $p(j_1), p(j_1, j_2), \ldots, p(j_1, j_2, \ldots, j_l)$, along with possibly other sequences generated over the time interval $[0,t]$. 

We let $\PP(*)$ denote the probability of event $*$ and $\EE[**]$ denote the expectation of random variable $**$.
Let $T_\rho(i)$ be the expected time for a product generated by entity  $i$ to percolate (within $D_\G$) to every entity in $\G$, and let $T_\rho = \max\{T_\rho(i): i \in G\}$. For  a wide range of standard percolation processes, the following properties then hold: (i) for $\rho>0$, we have $\EE[T_\rho]< \infty$;  (ii) for all $\eta>0$, $\lim_{\rho \rightarrow 0} \PP(T_\rho(i) >\eta) = 1$, and (iii) $\lim_{\rho \rightarrow \infty} \EE[T_\rho]=0$.
This last property implies that when $\rho$ is large, products are highly likely to percolate throughout the entire group $\G$ in a short time.

If we start this process at time $0$ with no products present,  let  $\tau_1, \tau_2, \ldots,\tau_k$ be the random variables that describe the time intervals between the generation of products across the collection of entities in $\G$.  By the assumptions of the model, the $\tau_i$  variables are independent and exponentially distributed, with each variable having an expected value of $1/(N\lambda)$. Thus, $\sum_{i=1}^k \tau_i$ is the time until $k$ products have been generated (this has a gamma distribution with expected value $k/(N\lambda)$). 
Let $\mu= N\lambda$. Then, for any $\eta>0$,  $\PP\left(\bigcap_{i=1}^k \{\tau_i \geq \eta\}\right)  = e^{-\mu k\eta}$ and $\PP(T_\rho \leq  \eta) \geq 1-\EE[T_\rho]/\eta$ (by the Markov inequality).   Let $\cE_k$ denote the following event: for  each of the first $k$ products generated, each product percolates to each entity in $\G$ before the next new product (in this collection of size $k$) is generated in $\G$. 
We then have:
\begin{equation}
\label{ppeq}
\PP(\cE_k) \geq  e^{-\mu k\eta} \cdot (1-\EE[T_\rho]/\eta)^k = (e^{-\mu\eta}(1-\EE[T_\rho]/\eta))^k.
\end{equation}
Setting $\eta = \sqrt{\EE[T_\rho]}$ in (\ref{ppeq}) and applying Property (iii) above gives:
$$\lim_{\rho \rightarrow \infty} \PP(\cE_k) =1.$$
Thus, as $\rho$ becomes large, the entities evolve collectively, and any variation is transient and short-lived. We will refer to this limiting case  as the {\em  community-based model}. 
One can model this  process by the following novel type of P{\'o}lya Urn model:

\begin{quote}
Consider an urn that initially has a single white ball.
At each step (the timing of which follows a Poisson process at rate $r$), a ball is selected from the urn uniformly at random.
If the selected ball is white, it is returned to the urn along with a ball of a new colour (not present already in the urn).
If the selected ball is coloured, it is removed and replaced by a ball of the same colour but a darker shade.
\end{quote}

To connect this urn process to the community-based model described above, note that selecting a white ball corresponds to the generation of a new product (which results in a ball of a new colour being added to the urn), while selecting a coloured ball and exchanging it for a darker one of that colour corresponds  to the transformation of an existing product. Thus, $r=N\lambda$.

We now compare the community-based model (corresponding to $\rho$ large) to the opposite extreme, where $\rho$ becomes small. In that latter case, 
 the probability that there is percolation between any two entities in $\G$ over the interval $[0, t]$ tends to $0$, and so products are only generated within entities but not shared between them.  We will refer to this limiting case as the {\em individual-based  model}. 
Note that in this individual-based model,  entity $i_j$ may possibly generate a new product $p(i_j)$,  or generate $p(i_j)$  and then transform it  (producing $p(i_j, i_j)$ and so on (or it might not generate any new products at all). Note that, in general, $p(i_j)$ may be different from $p(i_k)$ (for $k \neq j$) (i.e., different entities may either produce or transform different products). 

For the individual-based model, we have $N$ independent samples of the above Urn model but with $r=\lambda$.  By contrast, with the community-based model, we have a single sample of the above Urn model, but with $r=N\lambda$. 
Note that both  models have the same expected number of generation events, but they have quite different dynamics, as we now describe.  

Firstly, in the community-based model, there is only short-lived or transient variation among  the entities, whereas in the individual-based model, the individuals diverge from each other in terms of the collections of products available to them.
However, a subtler difference is that in the community model, the complexity of items is significantly higher than in the individual model, in a sense that we now make precise.

To analyse this in more detail, let $X_t$ denote the number of steps in this Urn process (i.e., where a ball is sampled and the urn modified) over the interval $[0, t)$.  Then $X_t$ has a Poisson distribution with mean $rt$.
Next, let $Y_t$   denote the number of times a white ball  is selected from the urn over the interval $[0,t)$, let $Z_t$ denote the number of times a coloured ball is selected from the urn over the time interval $[0,t)$, and let $C_t$ denote the number of coloured balls in the urn at time $t$.  Notice that the following two identities hold:
\begin{equation}
\label{eq1}
X_t = Y_t +Z_t, 
\end{equation}
\begin{equation}
\label{eq2}
Y_t = C_t = \mbox{the number of balls in the urn at time  $t$ minus 1}.
\end{equation}

Let $k_r:=\lfloor 2\sqrt{rt}\rfloor$. We claim that as $r \rightarrow \infty$:
\begin{equation}
\label{eq3}
\PP\left(X_t \geq \frac{1}{2}rt\right) \rightarrow 1 \mbox{ and } \PP(Y_t \leq k_r) \rightarrow 1,
\end{equation}
and so (applying the Bonferroni inequality):
\begin{equation}
\label{eq4}
\PP\left(X_t/Y_t \geq \frac{1}{2}rt/k_r\right)  \rightarrow 1.
\end{equation}
The first limit in (\ref{eq3}) holds because $X_t$ has a Poisson distribution with mean $rt$, and so $X_t/rt$ converges in probability to 1.  
For the second limit in (\ref{eq3}), let $T_{k_r}$ denote the time until $Y_t$ first hits $k_r$, in which case, a well known identity applies:
\begin{equation}
\label{eq5}
\PP(Y_t \leq k_r) = \PP(T_{k_r} \geq t).
\end{equation}
Now, $T_{k_r}$ is a sum of $k_r$ independent exponential random variables, with means $1/r, 2/r, \ldots, k_r/r$ and variances $1^2/r^2, 2^2/r^2, \ldots, k_r^2/r^2$.
Thus, $T_{k_r}$ has expected value  $$\EE[T_{k_r}] \sim k_r^2/2r \sim  4rt/2r = 2t,$$  and variance $$\VV ar[T_{k_r}] \sim k_r^3/3r^2 = 8 r^{3/2}t^{3/2}/3r^2 = \Theta(r^{-1/2}) \rightarrow 0 \mbox{ as } r \rightarrow \infty.$$
Applying  Eqn.~(\ref{eq5}) now shows that $\PP(Y_t \leq k_r) \rightarrow 1$ as $r$ grows, thereby justifying
the second part of Eqn.~(\ref{eq3}).

We now apply Eqns.~(\ref{eq1}) and (\ref{eq2}). These reveal that the condition $X_t/Y_t  \geq k_r$ is equivalent to the condition that $Z_t/C_t \geq k_r-1$. By the well-known `pigeonhole principle' in combinatorics, this last inequality implies that at least one of the ($C_t$) coloured balls must be at least $k_r-2$ shades darker than when that colour first appeared in the urn (because otherwise, each of the $C_t$ coloured balls must be selected at most $k_r-2$ times, in which case $Z_t \leq (k_r-2)C_t$).

It now follows from Eqn.~(\ref{eq4})  (noting that the term in  that equation (namely $\frac{1}{2}rt/k_r$) is of order  $\sqrt{r})$), that 
for large $r$ (for fixed $t$), there is a high probability that at least one coloured ball is present in the urn that is of order $\sqrt{r}$ shades darker than it was when it first appeared in the urn. Since the community-based model has $r=N\lambda$, we arrive at the following conclusion regarding the influence of the size of $\G$ on  complexity:

\begin{quote}
Over a  given period of time, some products in the community-based model  have an expected complexity of order  at least $\sqrt{N}$.
\end{quote}

By contrast, for the individual-based model, we have  $r=\lambda$ for each entity, and so we have $N$  independent  and identically distributed samples of a process where the maximum complexity of products across each group $\G$ exhibit a lower (logarithmic) dependence on $N$. (Moreover, these complex products are likely to exist only in one or a few entities, rather than being shared across the group).  To see this, note that the complexity of any product associated with an entity (up to time $t$) is bounded above by the number of generation steps for that entity, which  has a Poisson distribution with mean $\lambda t$, and the maximum of $N$ independent and identically distributed  Poisson random variables is known to be dominated asymptotically (with $N$) by a $\log(N)$ term \cite{kimber}. 

We note that the above model is different from the model described by \cite{vir15}), which was based on autocatalytic `cores' (which are closely related to RAFs \cite{hor16}) and also involved an ambient food set. However, the goal in that paper was to establish bounds on the rate of growth in the number of  autocatalytic cores as a function of the number of molecular species in a chemical reaction system.

\bigskip 

\section{Self--Other Reorganisation (SOR)}
\label{sorsec}

We have shown how entities that possess an abstract structure mathematically described by RAFs can exhibit evolution---i.e., cumulative, adaptive change---and how it is possible for this process to operate in the absence of variation and/or selection.
We refer to this kind of evolutionary process as {\em Self--Other Reorganisation} (SOR) because it consists of (i) self-organization of RAF networks, and (ii) interactions amongst these networks that alter their potential for future configurations \cite{gab13,gab19b,gabsmi18,vooetal20}. SOR \blue{was developed on the basis} of theory and findings that have come out of the post-modern synthesis in biology \cite{gab06,smietal18}. We now compare SOR to other established concepts in evolutionary theory: learning, natural selection, Lamarckian evolution, and the neutral theory of molecular evolution.

\subsection{SOR versus Learning}
Learning plays a role in SOR, since RAFs are central to SOR, and in cognitive RAFs the foodset  consists of MRs acquired through social learning and individual learning. These processes constitute the external component of SOR, as opposed to the internal component, which involves catalyzed reactions amongst MRs resulting in new MRs that were {\it not} available as initial reactants, i.e., foodset-derived MRs.
In the extreme, SOR could work with just one entity. In this case, the foodset consists of MRs acquired through  individual learning; there are no MRs acquired through social learning. However, so long as individually learnt information can trigger cognitive `reactions', it would be possible for some degree of cumulative, adaptive change to take place. There is a long history in psychology of viewing the assimilation of new information and the accommodation of existing knowledge in response to this new information as working hand in hand. Assimilation---the learning component---adds foodset itmes, but accommodation---an internal process---adds foodset-derived items. Thus, to the extent that they do work hand-in-hand, learning can be considered a kind of SOR. However, SOR is a broader concept than learning, because in the general case it operates on groups of entities, and can span biological generations.
It is unlikely that the converse is true, i.e., that evolution by SOR could work through social learning alone, with no individual learning, because there would be nothing to socially transmit. As demonstrated in a computational model of cultural evolution, turning off social learning resulted in cumulative adaptive change (exclusively within individual agents), albeit much more slowly. However, if social learning was the {\it only} means of learning available, there was nothing to kickstart the process, so there was no cumulative adaptive change \citep{gab95}.

\subsection{SOR versus Natural Selection}

\begin{figure}
\centering
\includegraphics[width=1\textwidth]{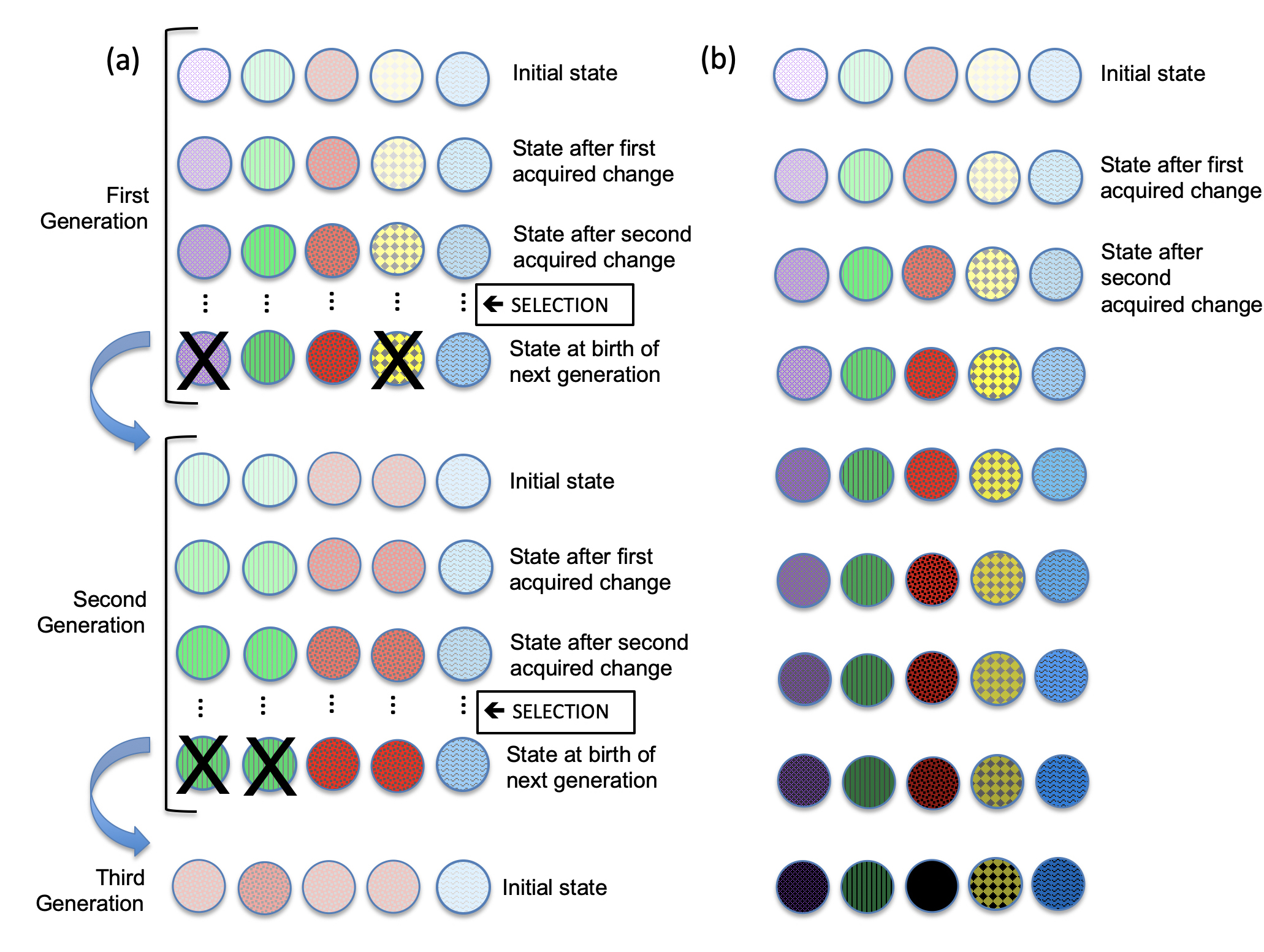}
\caption{Comparison of (a) evolution through  natural selection and  (b) evolution by Self--Other Reorganisation (SOR). An X indicates that the entity was selected against (i.e., did not contribute to the next generation). Individual entities are represented by circles, with variation among them represented by different patterns. (In the example shown in (b), there is variation but there need not be, as illustrated in Fig.~\ref{fig1}). In both (a) and (b), the darker the color the better adapted the individual (regardless of pattern). In (a), the darker color of one of the individuals in the third generation is due to mutation. Both (a) and (b) result in cumulative, adaptive change over time. However, (a) is relatively slow because it works due to differential replication of randomly generated heritable variation in a population over generations, such that some traits become more prevalent than others, with acquired changes being discarded at the end of each generation. In contrast, change in (b) is entirely due to acquired change over time.
}
\label{fig:sor}
\end{figure}

SOR is distinctly different from a Darwinian or selectionist process. The distinction between Darwinian evolution and SOR is summarised in Table~\ref{tab:sor} and illustrated in Fig.~\ref{fig:sor}.
SOR involves not \emph{competition and survival of some at the expense of others}, but \emph{transformation of all}.\footnote{Although SOR is an adaptive evolutionary process, communal exchange is not necessarily beneficial to the recipient. Transmission of useful plasmids through horizontal exchange among bacteria or protists may be beneficial, but transmission of viruses may be damaging. Similarly, transmission of useful technologies may be beneficial to the recipient, but transmission of misinformation may be harmful.} In other words, it isn't that the fittest survive (in their entirety) while the less fit do not, but rather, all agents develop more adaptive structure (such as happens in learning and creative cognition).\footnote{It has been suggested that creative cognition occurs through a Darwinian process \cite{cam74}, but this theory has been sharply criticized \cite{das11,gab07,gab19,ste98}, and its current primary proponent now advocates a non-Darwinian version of the theory \cite{sim10}.}
Like natural selection, SOR has mechanisms for preserving continuity and introducing novelty, but reproduction is lower-fidelity than it is for natural selection, because it is the culmination of haphazard catalyzed interactions, as opposed to the accuracy afforded by copying from a self-assembly code, as in a Darwinian process. 

\begin{table}[ht]
\begin{center}
\resizebox{15cm}{!}{
\begin{tabular}{@{} lll @{}}
\hline \hline 
Feature & \textbf{Variation--Selection} & \textbf{Self--Other Reorganisation (SOR)}\\ 
\hline
Unit of self-replication\tablefootnote{In SOR there is no replication by way of a self-assembly code. However, there is replication of a more piecemeal, imprecise kind that occurs through the transmission/duplication of components. 
\blue{The replication of conceptual RAF structure (e.g., when a parent teaches a child how to cook) may entail the generation of novelty (e.g., the child may intentionally double the quantity of chocolate chips), but replication has still taken place. Indeed, novelty gets introduced each generation even when organisms replicate; offspring are not identical to parents.}
We do not use the term `replicator' here since it is often assumed that replicator evolution necessarily involves inheritance of germ-line material.}
& Organism & RAF network \\
Preservation of continuity & Reproduction (vertical) & Communal exchange (horizontal) / social learning
\\
Generation of novelty &	Mutation, recombination	& Creative thinking, catalysis, transmission error \\
Self-assembly code 	& DNA or RNA & None \\
High fidelity &	Yes	& No \\
Transmission of acquired traits	& No & Yes \\
Type & Selectionist	& Lamarckian (by some standards; see above)\\
Evolution processes explained	& Biological &	Early life, horizontal gene transfer, culture \\
\hline \hline
\end{tabular}
}
\end{center}
\caption{Comparison between evolution through selection and evolution through Self--Other Reorganisation.}
\label{tab:sor}
\end{table}


A \emph{self-assembly code} is a set of coded instructions that is (i) \emph{actively interpreted} through a developmental process to generate a soma, and (ii) \emph{passively copied without interpretation} through a reproduction process to generate self-copies, i.e., new sets of self-assembly instructions that are in turn used in these two distinct ways \cite{hol92,lan92}.
This conception of a self-assembly code grew out of Von Neumann's work on universal constructors and relf-replicating automata \cite{von66}.
In a Darwinian  process, traits acquired through development of the soma and its interaction with the environment
(\emph{active interpretation} of the code) do not affect traits vertically transmitted through reproduction 
(\emph{passive uninterpreted} use of the code). The self-assembly code passed down through the germ-line is \emph{sequestered}, i.e., protected from the influence of acquired traits \cite{hol92}. Thus, the division of labour between these two ways of using the self-assembly code is responsible for the signature characteristic of a selectionist process: lack of transmission of acquired traits.
As a result, while acquired traits (those obtained during an organism’s lifetime, such as a tattoo, or knowledge of a recipe) may affect the soma, and potentially affect interpretation of the self-assembly instructions, they do not affect the \emph{content} of the self-assembly instructions.
Therefore, though inherited traits (e.g., blood type) get transmitted vertically from parent to offspring by way of genes, acquired traits do not. Thus, it is the division of labour between these two ways of using the self-assembly code that is responsible for the {\em sine qua non} of a Darwinian process: lack of transmission of acquired traits.

Now let us turn to the earliest structures that could be said to be alive, prior to the evolution of something as complex as a DNA- or RNA-based self-assembly code. Without a self-assembly code, there were no vertically inherited traits; all change was horizontally transmitted (i.e., acquired). Therefore, the evolution of early life was not due to differential replication of heritable variation in response to selection; it was non-Darwinian \cite{gab06,vet06,woe02}.

The situation is analogous for culture. Human culture does not (currently) involve a self-assembly code that gets used in these two distinct ways, and as such, it does not exhibit this signature trait of a selectionist process. No traits are inherited by way of a self-assembly code; all change is acquired, and acquired change is horizontally transmitted.\footnote{Note that vertical and horizontal transmission must be defined with respect to the relevant evolutionary process. Transmission of cultural information from parent to offspring is sometimes erroneously referred to as vertical transmission (e.g., \cite{cavfel81}). Although the individuals in question are in a parent--child relationship with respect to their status as biologically evolving organisms, this may not be the case with respect to their status as participants in cultural evolution. Indeed, although childbirth entails one mother and one father, there is no limit to the number of `parental influences’ on the `birth’ of an idea. 
A related error is to say that in cultural evolution, there is a third form of transmission--oblique transmission--in which traits are transmitted from non-kin members of the parental generation (e.g., \cite{cavfel81}). As far as cultural evolution is concerned, it is not strictly relevant whether the information comes from biological kin or non-kin. 
In a similar vein, although dual inheritance theorists speak of culture as a second form of inheritance \cite{henmce07,ricboy78,whi17,mul17}, the distinguishing feature of an inherited trait is that it is transmitted vertically (e.g., from parent to offspring) by way of a self-assembly code (e.g., DNA), and therefore not obliterated at the end of a generation. This is not the case with respect to cultural traits \cite{gab11} (nor is it even the case for all biological traits).} Although some elements of culture, such as kinship terminologies, have a generative logic, they do not use a self-assembly code: a set of coded instructions that is (i) actively interpreted through development to generate a soma, and (ii) passively copied without interpretation through reproduction to generate self-copies, i.e., new sets of self-assembly instructions that are in turn used in these two distinct ways \cite{vooetal20}.

One might also be tempted to suggest that natural language is a cultural self-assembly code. However, (1) although natural language involves \emph{encoding,} it is not a set of encoded instructions for the self-replication of natural languages, and (2) language evolution does not exhibit the signature characteristic of evolution by way of a self-assembly code: lack of transmission of acquired traits. Though some have argued that humans are biologically predisposed to achieve language, language itself is characterised by horizontal---not vertical---transmission. Therefore, language evolution is not due to the mechanism Darwin proposed: differential replication of heritable variation in response to selection \cite{gab04,gab13}. 
Results from computational modelling suggest that to cross the `Darwinian threshold' from non-selectionist to selectionist evolution requires the emergence of a self-assembly code \cite{vet06}. There is no evidence that language or any other component of culture has crossed this threshold, though it is not impossible that culture is moving toward a `cultural Darwinian threshold'; in other words, it may exist in the state biological life was in before the last universal common ancestor \cite{woe98}.

We posit that both early life and culture evolve not through a Darwinian process but through SOR.
A Darwinian explanation works in biology to the extent that retention of acquired change is negligible compared with retention of selected change; otherwise, the first (which can operate instantaneously) can overwhelm the second (which takes generations). Moreover, the period we associate with biological generations, from birth through to reproductive maturity and parenthood, is in general significantly longer than the stretch of time between when an individual acquires a cultural trait (e.g., an idea) and then expresses (their own version of, or their own take on) that cultural trait. That is, although {\it enculturation} is a lifelong process, a {\it particular item of cultural information} can be transmitted almost instantaneously, and hundreds may be transmitted back and forth during a conversation. This can make the relative speed of Darwinian evolution slower still.

\subsection{SOR versus Lamarckian Evolution}
As mentioned above, the term `Lamarckian' is widely understood to refer to the transmission of characteristics to offspring that were acquired through use or disuse during the entity's lifetime. 
According to interpretations of Lamarckism that specify that Lamarckian evolution requires genetic transmission to biological offspring (e.g., \cite{hul88}), SOR would not be classified as a form of Lamarckian evolution, as it is not restricted to biologically evolving entities. According to less restrictive interpretations of Lamarckism that encompass cultural as well as biological change (e.g., \cite{mes04}), SOR could be considered a form of Lamarckian evolution. 
However, unlike Lamarckism, SOR is precise about the kind of structure required (specifically, SOR requires RAF networks), and why this mechanism makes cumulative, adaptive change possible (as described above).


\subsection{SOR versus the neutral theory of molecular evolution}
The neutral theory of molecular evolution posits that much genetic variation in populations is the result of mutation and genetic drift, not selection \cite{kim68,kin69}. 
Random genetic drift in geographically isolated populations can lead to new forms appearing and others disappearing, without selection. Neutral change and drift are also observed in cultural evolution \cite{hah03,rucetal17}.
However, drift is as likely to be disadvantageous as it is to be advantageous---it is (statistically) neutral.
SOR, in contrast to neutral evolution, is not (statistically) neutral.
Like natural selection, SOR is a mechanism of cumulative, adaptive change.

\section{Implications}

The feasibility of evolution in the absence of variation and selection ---i.e., evolution by SOR---and the claim that early life and cultural evolution are both promising candidates for this second form of evolution, imply that we must be cautious about applying concepts and methodologies developed in a Darwinian evolutionary context in these two domains.
Since biological acquired traits are usually (though not always) discarded, and since a self-assembly code must stay intact to preserve its self-replication capacity, the joining of bifurcations in biological lineages is infrequent; thus, a phylogenetic tree correctly captures its branching structure. Speciation makes inter-lineage transfer of information relatively rare in biological organisms. 
By comparison, since cultural acquired traits are not discarded, and there is no cultural self-assembly code, the joining of bifurcations in cultural `lineages’ is commonplace, and thus cultural `lineages' tend to be network-like rather than tree-like \cite{gab06b,lip05,temeld07}. 
These distinctions become clear in phylogenetic studies. For multicellular species, phylogenetic networks calculated for different protein or RNA sequences tend to be tree-like.  However, phylogenetic trees constructed for different protein or RNA sequences across protists, bacteria or viruses reveal much more blending and greater network-like organisation \cite{shi16}. Gene expression profiles exhibit cross-similarity and greater convergent evolution as well.
Since cultural relatedness frequently arises through horizontal (inter-lineage) transmission, there is extensive blending of knowledge from different sources.  
 
Another (related) problem that arises when methods developed for selectionist evolutionary processes are applied to culture is due to convergent evolution, in which similar forms arise independently because they are alternative solutions within similar design constraints.  Examples include (i) the body shape and structure similarity between the Tasmanian tiger and the fox, (ii) wasp-imitating hover flies and (iii), the origin of basic and more complex brain structure across the tree of life \citep{leo, yos}. Because biological organisms must solve many problems (reproduction, locomotion, digestion, etc.), the probability that a species will be mis-categorised because of convergent evolution (i.e., on the basis of how it solves any one problem) is low. Cultural artifacts, on the other hand, are generally constructed with a single use in mind (though artifacts developed for use in one context may be used to solve other problems; for example, a screwdriver may be used to open a can of paint.) Thus, for cultural outputs, the probability of mis-categorisation arising through the assumption that similarity reflects homology is significantly higher. Therefore, the cost of proceeding as if a Darwinian framework were applicable to culture when it is not is high. 
Some have claimed that in practice this does not invalidate a phylogenetic approach to culture \citep{greetal09}. However, such conclusions come from analyses of datasets that involve little horizontal transmission (indeed, the creative blends that are the source of cultural novelty are often treated as `outliers' and are intentionally discarded from analysis).
This problem does not arise with SOR because it does not assume that superficially similar artifacts are homologous. Indeed, viewing culture from the perspective of SOR suggests that what is evolving is the structures of the conceptual networks that generate artifacts, not the artifacts themselves \citep{lan09}.

Such considerations have led some to develop network-based models of cultural evolution \citep{busetal19,endetal11,gab95,gabsmi18,gabetal11,kir17,lip05,vel12}. 
This body of research suggests that horizontal transmission can significantly alter the pattern of relatedness. For example, a network-based analysis of Baltic psaltery data that incorporated not just superficial physical attributes but also abstract conceptual attributes (such as markings indicative of sacred symbolic imagery), it was possible to resolve ambiguities arising from a phylogenetic analysis and generate a lineage more consistent with other historical data \citep{vel12}.
Horizontal cultural transmission may involve change in superficial features despite the preservation of deep structure,
as occurs in metaphor \cite{lak93}, analogy \citep{gen83,holtha96}, and cross-domain transfer, in which a source from one domain (e.g., music) inspires or influences a creative work in another (e.g., painting) \citep{ranetal13,scoetal19}. This kind of complexity and hierarchical structure cannot be captured without taking a network approach to cultural evolution, which provides further support for the theory that culture evolves through SOR.

Interestingly, similar issues arise with the simplest life forms. 
Because of phenomena such as mutualism, lineage reticulation (due to horizontal gene transfer and allopolyploidy---the combining the genomes of different parental species), certain traits evolve with astonishing speed, thereby diminishing the continuity and distinctiveness of species \cite{gogtow05,woe02}. Indeed, the stability of genetic information is so compromised that sequencing three {\em Escherichia coli} genomes revealed that fewer than 40\% of the genes were common to all three \cite{wel02}. As a result, the boundaries between many prokaryote species are fuzzy, and exhibit reticulate patterns of evolution, thus calling into question the appropriateness of the notion of the “tree of life” \cite{hilgog93,hug08, mci}. The limitations of Darwinism as an explanation of the forms and dynamics of living things is increasingly recognised, and the role of epigenetic processes has become increasingly appreciated. Nevertheless, because reticulate (horizontal) proceses are much less present in biological evolution than cultural evolution, natural selection provides a reasonable approximation.

\section{Concluding comments}

Although natural selection is a {\it theory} of evolution (a spectacularly successful one in biology), the term `evolution' is sometimes assumed to be synonymous with variation and selection.
Even in research on cultural evolution---which is sometimes (though not universally) regarded as Lamarckian---it is generally assumed that variation and selection are essential for culture to evolve. 
Using RAF networks, this paper demonstrates by way of example, or `existence proof' that evolution is possible in the absence of variation, selection, and competition. 
We refer to this kind of evolution as Self--Other Reorganisation (SOR) because it occurs not through competitive exclusion such that only the fittest reproduce, but through the assimilation, restructuring, and exchange of components. SOR is a primitive form of evolution that can generate change quickly since it does not require the discarding of acquired traits, and it can take place in the absence of birth or death. Because it does not use coded self-assembly instructions, it is more haphazard than natural or artificial selection, but sufficient for cumulative, adaptive change. 

RAFs have proven useful in two areas of research that might appear to be quite distinct, but that both involve networks and evolutionary processes. In the OOL, the underlying reaction system is a biochemical reaction network, and in the OOC it is a conceptual network.
Since cultural evolution lacks a self-assembly code, and lacks the signature characteristic of Darwinian evolution---discarding of acquired traits at the end of a generation---it seems reasonable to investigate the extent to which the OOL and the OOC share a common theoretical framework.
This is consistent with the proposal that the entities that evolve through culture are not discrete artifacts or units of behaviour such as songs, jokes or memes, but the minds that generate them \cite{gab04}, which have a structure that is self-organising and self-maintaining \cite{greetal02,matvar73}, and therefore lends itself to the application of RAF formalisms \citep{gabste,gabste20,gabste-modern}.
We suggest that SOR may be operative during the early stage of any evolutionary process, that it was responsible for both the origin and early evolution of both organic life and human culture, and that RAFs provide a means of modeling evolution through SOR. 
The approach offers an established formal framework for integrating biological, cultural, and cognitive proceses, and embedding this synthesis in the study of self-organizing structures and their role in evolutionary processes. 
\blue{The framework suggests avenues for future research, such as elaboration of the RAF model to include how new elements might be synthesized from the foodset elements and previous foodset-generated elements, and to describe how new elements might be secreted from one RAF and incorporated into another.
}

SOR is a broader concept than learning, as it can produce adaptive change in groups of entities, and it can span biological generations. Competition may occur in SOR, but unlike selectionist evolution it is not essential.
Because SOR does not require competition or variation, it may increase homogeneity among members of a culturally evolving group of individuals by increasing the amount of shared knowledge and experience among group members. It may thereby play an important role in fostering group identity, cohesion, and cooperation \cite{vooetal20} (see also \cite{andtor19}).

\newpage
\doublespace

\section{Supplementary Information: Alphabetical list of Definitions of Key Terms}

Since miscommunication arises from inconsistencies in how evolutionary concepts are applied in the sciences and social sciences, to maintain clarity we provide an alphabetical list of standard definitions of key terms used in this paper.

\begin{itemize}

\item {\bf Acquired trait:} a trait obtained during the lifetime of its bearer (e.g., a scar, tattoo, or the memory of a song) and transmitted horizontally (i.e., laterally). 

\item {\bf Culture:} extrasomatic adaptations—including behavior and artifacts—that are socially rather than sexually transmitted.

\item {\bf Darwinian (or `selectionist') process:} an evolutionary process---i.e., a process that exhibits cumulative, adaptive, open-ended change---occurring through natural or artificial selection.

\item {\bf Darwinian threshold:} transition from a non-Darwinian to a Darwinian evolutionary process \cite{woe02,vet06}.

\item {\bf Evolution:} descent with modification, or cumulative, adaptive change over time, giving rise to new species that share a common ancestor \cite{darwin}. The most well-known theories to explain how evolution works are natural selection and Lamarckism.
The concept of evolution has been extended to apply to, not just to cumulative, adaptive change in biological species, but also cumulative, adaptive {\it cultural} change \cite{boyric85,cavfel81}.
Thus, here we use the more general definition of evolution used in cultural evolution research: cumulative, adaptive
change over time.

\item {\bf Generation:} a single transition period from the internalized to the externalized form of a trait.\footnote{Note that, with respect to biological evolution, a new generation generally (though not in horizontal gene transfer) begins with the birth of one or more organism(s). With respect to cultural evolution, a `generation' may begin with the transmission of an idea (a cultural trait).
Thus, over the course of a discussion, the idea  may undergo multiple generations.}

\item {\bf Horizontal transmission:} non-germ-line transmission of an acquired trait from one entity to another. Thus, social transmission is horizontal because the information is not inherited, i.e., it is not passed from parent to offspring by way of genes.

\item {\bf Lamarckian process (or `soft inheritance'):} widely understood to mean the transmission of characteristics to offspring that were acquired through use or disuse during its lifetime.\footnote{We note that although this is how the term is standardly defined, it has been pointed out that this idea was not unique to Lamarck, and transmission of characteristics was only one component of Lamarck's theory \cite{gou02}.} Some maintain that Lamarckian evolution requires genetic transmission to biological offspring, a view held by early biologists (e.g., \cite{sem21}), though current scholars (e.g., \cite{mes04}) sometimes take a more equivocal view.

\item {\bf Neutral theory of molecular evolution}: the theory that a significant degree of genetic variation in populations is the result of mutation and genetic drift, not selection \cite{kim68,kin69}.

\item {\bf Selection:} differential replication of randomly generated heritable variation in a population over generations such that some traits become more prevalent than others. Selection may be natural (due to non-human elements of the environment) or artificial (due to human efforts such as selective breeding), and it can occur at multiple levels, e.g., genes, individuals, or groups \cite{lew70}.


\item {\bf Self-assembly code:} a set of coded instructions that is: (i) \emph{actively interpreted} through a developmental process to generate a soma, and (ii) \emph{passively copied without interpretation} through a reproduction process to generate self-copies, i.e., new sets of self-assembly instructions that are in turn used in these two distinct ways. \blue{(The word `instructions' should not be interpreted literally here as to imply that they are written in a natural language such as English. The instructions are implicit in the dynamics of the entity; it is structured in such a way as to ensure a sequence of processes that result in self-assembly.)}

\item {\bf Self--Other Reorganisation (SOR):} a theory of how both culture, and early life, evolve through interacting, self-organizing networks, based on theory and findings that have come out of the post-modern synthesis in biology  \cite{gab06,smietal18}.

\item {\bf Vertical transmission:} Germ-line inheritance of a trait from one generation to the next. (In other words, transmission occurs by way of a self-assembly code, such as DNA.)

\end{itemize}

\section{Acknowledgements} 
This research was funded by Grant 62R06523 from the Natural Sciences and Engineering Research Council of Canada.
We thank the editor and reviewers for helpful comments and suggestions on the manuscript.

\bibliographystyle{siamplain}
\bibliography{main2.bbl}

\end{document}